\documentclass[aps,prl,twocolumn,groupedaddress,showpacs]{revtex4-1}

\usepackage{graphicx}
\usepackage{changes}

\begin{document}
\title{High Sensitivity RF Spectroscopy of a Strongly-Interacting Fermi Gas}
\author{Constantine Shkedrov, Yanay Florshaim, Gal Ness, Andrey Gandman, Yoav Sagi}
\email[Electronic address: ]{yoavsagi@technion.ac.il}

\affiliation{Physics Department, Technion - Israel Institute of Technology, Haifa 32000, Israel}

\date{\today} \begin{abstract}

Rf spectroscopy is one of the most powerful probing techniques in the field of ultracold gases. We report on a novel rf spectroscopy scheme with which we can detect very weak signals of only a few atoms. Using this method, we extended the experimentally accessible photon-energies range by an order of magnitude compared to previous studies. We verify directly a universal property of fermions with short-range interactions which is a power-law scaling of the rf spectrum tail all the way up to the interaction scale. We also employ our technique to precisely measure the binding energy of Feshbach molecules in an extended range of magnetic fields. This data is used to extract a new calibration of the Feshbach resonance between the two lowest energy levels of $^{40}$K.
\end{abstract}

\maketitle

Degenerate Fermi gases have been used extensively in recent years to study complex many-body phenomena \cite{RevModPhys.80.885} such as high-$\mathrm{T}_c$ superfluidity \cite{Gaebler2010,Feld2012}, magnetic ordering \cite{Hulet2015Nature,GreinerNature2017antiferromagnet} and many-body localization \cite{Schreiberaaa7432}. Their usefulness lies in their universality and controllability, namely the ability to characterize their state with a relatively small number of intensive macroscopic parameters that can be externally tuned \cite{QunatumSimulationReview2012}. The prime example is the contact-like interactions between two fermions with opposite spins. Since the gas is dilute, the details of the inter-atomic interactions are usually not important and can be replaced by an effective contact pseudo-potential $V(\mathbf{r})=\frac{4 \pi \hbar^2 a}{m}\delta(\mathbf{r})\frac{\partial}{\partial r} r$, where $a$ is the s-wave scattering length, $m$ is the atomic mass and $\hbar$ is the reduced Planck's constant \cite{PhysRev.105.767}. The scattering length can be modified close to a Fano-Feshbach resonance, which occurs when the energy of a molecular bound-state is brought close to that of the colliding atoms, usually by controlling an external magnetic field \cite{RevModPhys.82.1225}. 

The contact-like interactions give rise to several universal relations derived by S. Tan and others, all tied-up through the contact parameter, $\mathrm{C}$ \cite{Tan08,*Tan08a,*Tan08b,Braaten08,PhysRevA.78.053606,PhysRevA.79.023601,Castin2009,Braaten2012}. Several of these relations were verified experimentally and the value of the contact was measured at different thermodynamic conditions \cite{PhysRevLett.95.020404,PhysRevLett.104.235301,Navon07052010,PhysRevLett.106.170402,PhysRevLett.109.220402,Hoinka2013}. This universal behavior is expected as long as the relevant scale is small compared to the interaction scale. For example, the spectral function has a characteristic back-bending at large energies and momenta which translates into power-law scaling of their respective distributions: $n(k)\sim \mathrm{C}/k^4$ for the momentum and $n(\epsilon)\sim \mathrm{C}/\epsilon^{3/2}$ for the energy \cite{Tan08,PhysRevA.81.021601,Braaten2012}. This scaling is expected to hold for $k<\frac{1}{r_0}$ and $\epsilon<\frac{\hbar}{m r_0^2}$, where $r_0$ is the effective range of the potential \cite{Tan08}. The probability to find an atom decreases rapidly in the tail of these distributions. Therefore, the range in which the universal scaling has been experimentally tested to date is more than an order of magnitude shorter than the interaction scale. In this letter, we address the challenge of reaching the interaction scale by developing a new high-sensitivity rf spectroscopy technique. Applying it to a degenerate Fermi gas around a Feshbach resonance, we establish a universal power-law scaling of the rf lineshape over more than two decades in frequency, up to the interaction scale. We demonstrate the universality of the power-law exponent for data taken at different interaction strengths. Finally, we use the enhanced sensitivity of our technique to precisely measure the Feshbach molecule binding-energy up to relatively high energies and obtain a new calibration for the $^{40}$K$_2$ Feshbach resonance position and width.

\begin{figure}
	\centering
	\includegraphics[width=1\linewidth]{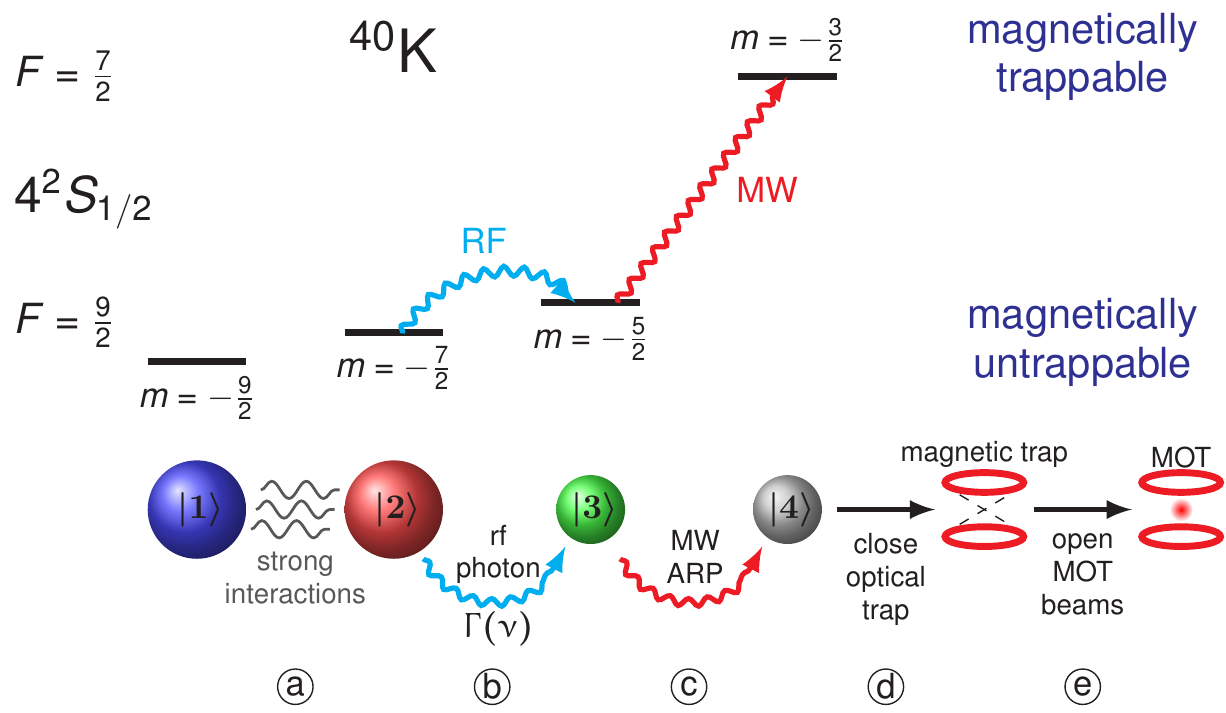}
	\caption{High sensitivity rf spectroscopy. A gas of $^{40}$K atoms in a balanced mixture of states $|1\rangle=|F=9/2,m=-9/2\rangle$ and $|2\rangle=|9/2,-7/2\rangle$ is prepared in $T/T_F\approx 0.2$ in the vicinity of a Feshbach resonance at $B=202.2$G (a). We then apply a $400\mu$s square pulse of rf radiation with a frequency $\nu$ close to 47MHz (b). This pulse transfers a small fraction of the atoms to state $|3\rangle=|9/2,-5/2\rangle$. Next, we transfer all the atoms in $|3\rangle$ to state $|4\rangle=|7/2,-3/2\rangle$ using MW sweep whose central frequency depends on the magnetic field but is typically around $1.6$GHz (c). Then the optical trap is turned off and a quadrupole magnetic field is ramped on (d). Due to the different signs of their magnetic dipole moments, atoms at $|1\rangle$,$|2\rangle$ and $|3\rangle$ are anti-trapped while the atoms at $|4\rangle$ are trapped by this field. After a waiting time that ensures only atoms at state $|4\rangle$ are present, the MOT laser beams are switched on (e). The fluorescence signal of the atoms captured in the MOT is recorded with a sensitive camera. The detection sensitivity can in principle reach a single atom, although in our apparatus it is limited to $\sim5$ atoms due to background scattering from surrounding optical surfaces.}
	\label{fig:the_experimetnal_sequence}
\end{figure}

The new rf spectroscopy method is depicted in Fig.~\ref{fig:the_experimetnal_sequence}. We prepare a quantum degenerate gas of $^{40}$K atoms in a balanced mixture of the two lowest Zeeman sublevels, denoted by $|1\rangle$ and $|2\rangle$, whose energy is split by a homogeneous magnetic field $B$. Rf radiation at a frequency $\nu$ transfers a small fraction of the atoms at state $|2\rangle$ to a third Zeeman sublevel $|3\rangle$, which is initially unoccupied. The number of atoms out-coupled by the rf pulse is recorded, and the result of this measurement is the rf transition rate $\Gamma(\nu)$. Many important observables can be extracted from $\Gamma(\nu)$, including pair size, superfluid gap, quasi-particle dispersion and single-particle spectral function, when the momentum distribution is resolved \cite{Chin1128,KetterleNature2008Determination,PhysRevLett.101.140403,PhysRevLett.102.230402,Nature_Stewart_2008,Gaebler2010,PhysRevLett.114.075301}. In standard rf spectroscopy, the number of the out-coupled atoms is measured by absorption imaging. Since the photon absorption cross-section is relatively small, detecting very weak signals in the rf lineshape tail is challenging. We overcame this difficulty by using fluorescence imaging instead. With fluorescence imaging, even a single atom can be reliably detected if held for a sufficient period of time, as was demonstrated with a magneto-optical trap (MOT) \cite{Hu:94}. For rf spectroscopy, the main challenge is to capture in a MOT \emph{only} the atoms out-coupled by the rf pulse, which may be a tiny fraction of the whole cloud. 

Our approach relies on the difference in energy and magnetic dipole moments of the various energy levels. Specifically, all three states $|1\rangle$,$|2\rangle$, and $|3\rangle$ have negative magnetic dipole moments (``high-field-seekers'') and are therefore magnetically untrappable \cite{WING1984181}. After the rf pulse is applied, we use a narrow microwave (MW) sweep to selectively transfer only the population at state $|3\rangle$ to a state $|4\rangle$ in the other hyperfine manifold that has a positive magnetic dipole moment and is therefore magnetically trappable (``low-field-seeker''). We then turn on a quadrupole magnetic field which magnetically traps the atoms in state $|4\rangle$ and expels all the rest. Finally, we turn on the MOT laser beams and record the MOT fluorescence signal with a sensitive CMOS camera.

The experiments are performed in a newly built apparatus that we now briefly describe. Our system is composed of three interconnected vacuum chambers. In the first chamber, a 2D MOT \cite{dieckmann1998two} generates a stream of cold atoms that fly through a narrow nozzle to the second chamber. There, the atoms are captured and cooled in a dark SPOT 3D MOT \cite{ketterle1993high} followed by a gray molasses cooling on the D1 line \cite{salomon2014gray}. The atoms are then optically pumped to a mixture of the $F=9/2,m_F=9/2,7/2$ Zeeman states, loaded into a QUIC magnetic trap \cite{esslinger1998bose} and cooled by forced MW evaporation down to $T/T_F\approx5$, where $T_F$ is the Fermi temperature. At this point the atoms are loaded into a far-off-resonance optical trap at a wavelength of $\lambda=1064nm$ \cite{Grimm200095}, undergo a short stage of forced optical evaporation and moved to the third chamber by moving the trap position a distance of $328$mm during $\sim 1$ second. After the transfer is complete, a second laser beam crossing the first beam at an angle of $45^\circ$ is turned on. We also apply rf radiation at $5$MHz and adiabatically sweep the magnetic field in order to transfer the atoms to negative spin states $m_F=-9/2,-7/2$ which have a broad Feshbach resonance around $B_0=202.2$G. The final stage of optical evaporation ends at a magnetic field of $203.4$G. At this stage, there are around $150,000$ atoms per spin state at $T/T_F\approx0.2$. The field is then ramped adiabatically to the final value $B$ and the system is left for $10$ms to fully equilibrate. The trap oscillation frequencies in most of the experiments are $2\pi\times 289$ Hz and $2\pi\times 24$ Hz in the radial and axial directions, respectively. The average trap Fermi energy is $E_F=h\times 12$kHz with $h=2\pi\hbar$. 

We probe the system using a $400\mu$s-long square rf pulse with a spectral resolution of $2.2$kHz. The rf power is chosen in the linear regime where it transfers less than $10\%$ of the atoms, and it is kept constant for all rf frequencies. Just before releasing the optical trap, we apply a $1.3$ms pulse of MW radiation and sweep its frequency across $200$kHz to induce adiabatic rapid passage (ARP) transfer to state $|4\rangle$. After the ARP, the optical trap is turned off abruptly and the quadrupole field is ramped on simultaneously, trapping the atoms at state $|4\rangle$. We then wait $100$ms for atoms in all other states to completely leave the MOT capture domain and switch on the MOT beams. The photons scattered by the atoms in the MOT are collected for $200$ms. Finally, we release the atoms from the MOT, wait for $400$ms, and repeat the detection sequence to obtain the background signal, which we subtract.

\begin{figure}
	\centering
	\includegraphics[width=1\linewidth]{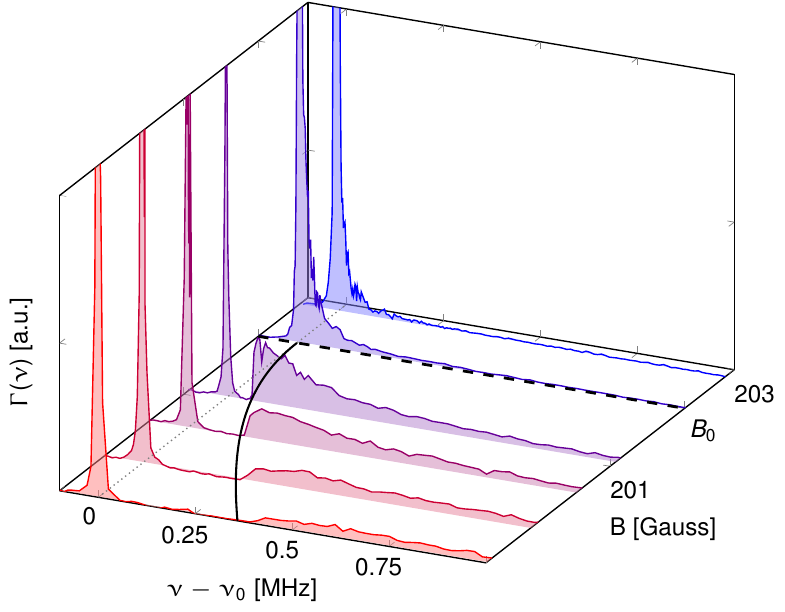}
	\caption{Rf transition rate, $\Gamma(\nu)$, taken at different magnetic fields in the vicinity of the Feshbach resonance at $B_0=202.2$G (dashed line). The frequency axis is plotted relative to $\nu_0$ - the transition frequency for non-interacting atoms, which is measured with a spin polarized gas. For clarity, only a range of 1MHz is shown here. The solid line is the theoretical Feshbach molecule binding energy given by Eq.(\ref{Eq_Eb_theoretical}) using $a(B)$ with the calibration reported in Ref. \cite{Gaebler2010}.}
	\label{fig:raw_data_at_different_fields}
\end{figure}

In Fig.~\ref{fig:raw_data_at_different_fields} we plot rf spectra taken at different magnetic fields close to the Feshbach resonance. A tail at high frequencies is apparent and its weight increases as the field is decreased. For fields $B<B_0$ the spectrum separates into two distinct features: a narrow peak at zero frequency owing to free atoms and a broad spectrum with a sharp onset owing to bound Feshbach molecules. The rf lineshape is normalized by $\int_{-\infty}^{\infty} \Gamma(\nu) d\nu=1/2$. For this choice of normalization and when the third spin state interacts only weakly with the two other states, as is the case with $^{40}$K, contact interactions in 3D give rise to a power-law scaling of $\Gamma(\nu)$ at high frequencies \cite{PhysRevA.81.021601,PhysRevLett.104.223004,Braaten2012}:
\begin{equation}\label{Eq_Gamma_vs_nu}
\Gamma(\nu)\rightarrow \frac{\mathrm{C}}{2^{3/2} \pi^2} \cdot \nu^{-3/2}\ \ ,
\end{equation}
where $\nu$ is in units of $E_F/h$ and $C$ is in units of $N k_F$, where $k_F$ is the Fermi wave-vector and $N$ is the total number of atoms. The increase of the spectral weight in the tail as $B$ is reduced is expected since the contact increases monotonically \cite{Braaten2012}.

\begin{figure}
	\centering
	\includegraphics[width=1\linewidth]{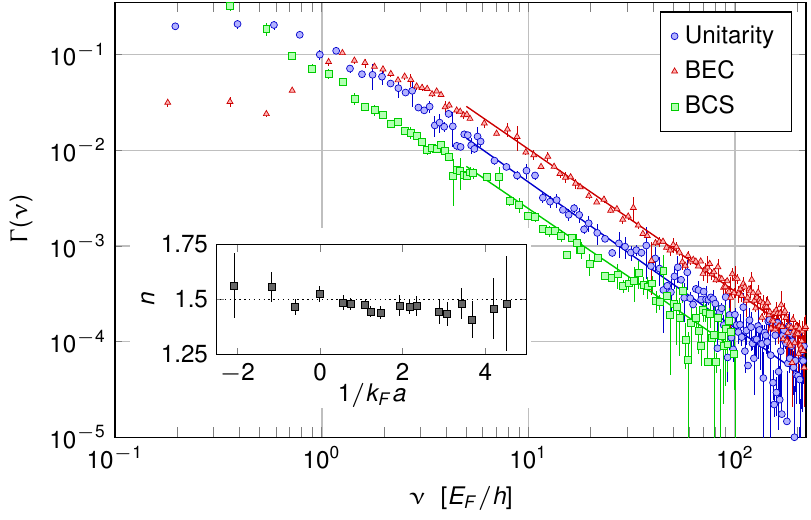}
	\caption{Rf lineshapes for three different interaction strengths in the BCS-BEC crossover \cite{BCS-BEC_Zwerger_book}: $1/k_Fa=0$ (unitarity), $1/k_F a=0.49$ (BEC) and $1/k_F a=-0.53$ (BCS). Linear scaling shows that the data follows a power-law at high frequencies. The inset shows the power-law exponent extracted by fitting the tail of each dataset with $c_1/\nu^n$, starting from $\nu>5E_F/h$ for $B>B_0$ and $\nu>E_b+20E_F/h$ for $B<B_0$, where $E_b$ is given by Eq.(\ref{Eq_Eb_theoretical}). The error bars correspond to one $\sigma$ confidence interval of the fit parameter.}
	\label{fig:logarithmic_plot_and_exponent}
\end{figure}

We first focus on the universal scaling of the rf lineshape. Previous studies at unitarity ($1/k_Fa=0$) with standard rf spectroscopy were restricted to $\nu<12E_F$ due to signal to noise ratio  \cite{PhysRevLett.104.235301,PhysRevLett.109.220402}. On the other hand, it was found empirically that the scaling of Eq.(\ref{Eq_Gamma_vs_nu}) starts to show up only above $5E_F$ \cite{PhysRevLett.104.235301}. In this limited range $5E_F<\nu<12E_F$, $\Gamma(\nu)$ decreases by less than a factor of $4$, hence only consistency with a power-law scaling could have been established \cite{PhysRevLett.104.235301,PhysRevLett.109.220402}. Using our new detection scheme, we are able to extend this range by more than an order of magnitude, as can be seen in Fig.~\ref{fig:logarithmic_plot_and_exponent}, where we plot on a logarithmic scale data taken at $1/k_Fa=0.49,0,-0.53$ in the BCS-BEC crossover regime \cite{BCS-BEC_Zwerger_book}. One of the main results of this work is the observation of linear scaling of the rf spectrum over more than two decades up to $150E_F-200E_F$, which establishes without any prior assumption that the rf lineshape tail indeed follows a power-law. Furthermore, we have taken rf spectra at many other interaction strengths and used a power-law function $\Gamma(\nu)\propto \nu^{-n}$ to fit the tail. The exponent $n$ extracted from these fits is plotted in the inset of Fig.~\ref{fig:logarithmic_plot_and_exponent}. The results are consistent with a value $1.5$ predicted by Eq.(\ref{Eq_Gamma_vs_nu}) over the range of interactions we have tested. For $^{40}$K, the interaction energy scale, $\hbar^2/mr_0$, is $\sim h\times 2.7$MHz or $\sim 225 E_F$. Our data extends to this scale and confirms the universal scaling resulting from contact-like interactions in the vicinity of the Feshbach resonance. 

Next, we study the Feshbach molecule spectra in Fig.~\ref{fig:raw_data_at_different_fields} which exist for positive scattering length ($B<B_0$). The binding energy of the Feshbach molecule is related to the scattering length through \cite{PhysRevA.48.546}:
\begin{equation}\label{Eq_Eb_theoretical}
E_b=\frac{\hbar^2}{m(a-\bar{a})^2} \ \ ,
\end{equation}
where $\bar{a}=\frac{1}{\sqrt{8}}\frac{\Gamma(3/4)}{\Gamma(5/4)}\left[m C_6/\hbar^2\right]^\frac{1}{4}$ is the finite range correction of the van der Waals potential $U(r)\approx-C_6/r^6$. Near a Feshbach resonance the scattering length can be written as $a(B)=a_{bg}\left(1-\frac{\Delta}{B-B_0}\right)$, where $\Delta$ is the width of the resonance and $a_{bg}$ is the background scattering length \cite{RevModPhys.82.1225}. A precise measurement of $E_b$ is valuable in order to calibrate the molecular potentials and parameters of the Feshbach resonance \cite{RegalNature2003molecules,PhysRevA.67.060701,GaeblerPhdthesis2010}.

We extract the binding energy from the position of the sharp rise in the molecular spectra, which marks the molecule dissociation threshold. A general form of the transition lineshape of a weakly bound molecules is 
\begin{equation}\label{Eq_molecule_lineshape}
\Gamma(\nu)=\Theta(\nu-E_b/h)\frac{\mathrm{C}}{2^{3/2} \pi^2}\frac{\sqrt{\nu-E_b/h}}{(\nu-\nu_w)^2} \ \ ,
\end{equation} 
where $\Theta$ is the Heaviside step function, and in the two-body limit $\nu_w\rightarrow 0$ \cite{PhysRevA.71.012713,KetterleNature2008Determination}. Note that Eq.(\ref{Eq_molecule_lineshape}) has the correct high frequency limit given by Eq.(\ref{Eq_Gamma_vs_nu}). As can be seen in Fig.~\ref{fig:raw_data_at_different_fields}, the molecular signal decreases as the magnetic field is taken farther away from the resonance. In this limit, the contact scales as $\mathrm{C}\sim 1/a$ \cite{Braaten2012} while the binding energy as $E_b\sim 1/a^2$. The maximum of the molecular lineshape in Eq.(\ref{Eq_molecule_lineshape}) scales as $\mathrm{Max}[\Gamma(\nu)]\sim \mathrm{C}/E_b^{3/2}\sim a^2$. Farther away from resonance $a$ falls off and consequently also the rf signal. Therefore, our sensitive rf spectroscopy technique is particularly well-suited to measure these weak and extended signals.

\begin{figure}
	\centering
	\includegraphics[width=1\linewidth]{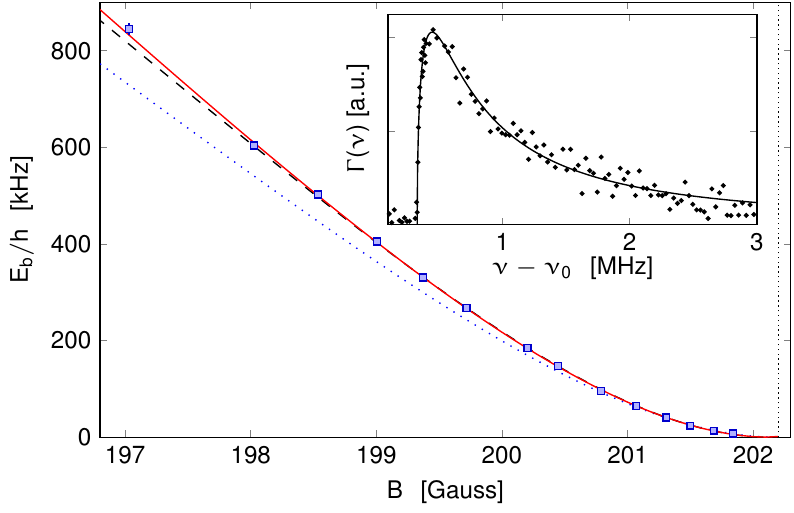}
	\caption{The binding energy (squares) of the Feshbach molecule at different magnetic fields close to the Feshbach resonance (dashed vertical line). We extract the binding energy by fitting the rf lineshape with the molecular spectral function given by Eq.(\ref{Eq_molecule_lineshape}). The inset shows a typical dataset taken at $B=199.371(8)$ and the fit (solid line), which yields $E_b/h=331(2)$ kHz with $1\sigma$ confidence level. The theory of Eq.(\ref{Eq_Eb_theoretical}) (dotted blue line) with the calibration of the Feshbach resonance parameters given in Ref. \cite{Gaebler2010,GaeblerPhdthesis2010} ($\Delta=7.04(10)$G, $B_0=202.20(2)$G and $a_{bg}=174 a_0$) shows an increasing systematic deviation from the data farther away from the resonance. We obtain a satisfactory fit to the data with a two coupled channels calculation based on the model of Ref. \cite{0953-4075-37-17-006,PhysRevA.72.013610} with a new calibration: $B_0=202.14(1)$G, $\Delta=6.70(3)$G, and using the known values and $a_{bg}=169.7a_0$\cite{PhysRevA.78.012503}, $C_6=3897$ a.u. \cite{PhysRevLett.82.3589} and $\delta\mu/h=2.35$MHz/G \cite{PhysRevA.72.013610} (solid red line). Eq.(\ref{Eq_Eb_theoretical}) with the new calibration also fits the data well (dashed black line).}
	\label{fig:Eb_vs_nu}
\end{figure}

Typical experimental data together with a fit to Eq.(\ref{Eq_molecule_lineshape}) are shown in inset of Fig.~\ref{fig:Eb_vs_nu}. From these fits at different magnetic fields we determine $E_b(B)$ with an average relative accuracy of $2.5\%$ at a $1\sigma$ confidence level. The results are shown in Fig.~\ref{fig:Eb_vs_nu}. For comparison, we also plot the formula of Eq.(\ref{Eq_Eb_theoretical}) with the most recent calibration of the Feshbach resonance parameters \cite{Gaebler2010,GaeblerPhdthesis2010}. It fits well the data near the resonance, but for magnetic fields larger than $1$G a clear systematic deviation develops. A similar trend was also observed in Ref. \cite{PhysRevLett.94.210401}. In order to extract a new calibration, we calculate the binding energy using a two coupled channels model \cite{0953-4075-37-17-006,PhysRevA.72.013610}. This model depends on five parameters: $a_{bg},\Delta, B_0,C_6$ and $\delta\mu$, which is the difference in the magnetic dipole moment of the open and closed channels. We have fixed the values of $C_6=3897$ atomic units \cite{PhysRevLett.82.3589}, $\delta\mu/h=2.35$MHz/G \cite{PhysRevA.72.013610} and $a_{bg}=169.7a_0$ with $a_0$ being the Bohr radius \cite{PhysRevA.78.012503}, which have been determined to a good accuracy, and used $\Delta$ and $B_0$ as fitting parameters. This yields $B_0=202.14(1)$G, $\Delta=6.70(3)$G. As can be seen in Fig.~\ref{fig:Eb_vs_nu}, with this calibration both the two channels model and Eq.(\ref{Eq_Eb_theoretical}) fit the data very well. Our new calibration shifts $B_0$ and $\Delta$ slightly relative to their values in Ref. \cite{Gaebler2010,GaeblerPhdthesis2010} and improves their accuracy by a factor of $2$ and $3.3$, respectively.



In conclusion, we have presented a new sensitive detection scheme for rf spectroscopy which improves this already powerful technique. The new scheme allows detection of very weak signals of only several atoms. The new technique has been employed to confirm the universal behavior of a contact-like potential all the way to the microscopic interaction scale. In addition, the Feshbach molecule binding energy has been measured in an extended range of magnetic fields and used to extract a new calibration of the Feshbach resonance parameters $B_0$ and $\Delta$. We expect the new technique to be useful in many applications where the signal is inherently weak, e.g. when a precise determination of a minority species spectrum is of interest. In addition, it would be interesting to measure the universal scaling of the rf lineshape in lower dimensions. In particular, in 2D one expects a power-law with a different exponent of $2$ \cite{PhysRevLett.108.060402}. Moreover, crossover from 2D scaling to 3D scaling is expected at an energy scale corresponding to $\hbar \omega_z$, with $\omega_z$ being the oscillation frequency in the tight direction of the 2D trap.

We thank John Bohn for insightful comments. This research was supported by the Israel Science Foundation (ISF) grant No. 888418, and by the United States-Israel Binational Science Foundation (BSF), Jerusalem, Israel, grant No. 2014386.

C. S. and Y. F. contributed equally to this work.

%
\end{document}